%% file: main.tex
\documentclass[journal]{IEEEtran}

\IEEEoverridecommandlockouts

\usepackage[cmex10]{amsmath}
\usepackage{siunitx}
\usepackage{cite}
\usepackage{psfrag}
\usepackage[utf8]{inputenc}
\usepackage[T1]{fontenc}
\usepackage{amsmath,amsfonts,amsbsy,amssymb}
\usepackage{mathabx}
\usepackage{mathrsfs}
\usepackage[nolist]{acronym}
\usepackage{tabularx}
\usepackage{amssymb}
\usepackage{amsmath}
\usepackage{graphicx}
\usepackage{cite}
\usepackage{multirow}
\usepackage{wasysym}
\usepackage{multirow}
\usepackage{float}
\usepackage{xcolor}
\usepackage{subcaption}
\usepackage{algorithm}
\usepackage{algorithmic}
\usepackage{xcolor}
\usepackage{suffix}
\usepackage{hyperref}
\usepackage{breakurl}
\usepackage{url}
\usepackage{hyperref}
\usepackage[font=footnotesize]{caption}
\usepackage{epsfig}
\usepackage[nolist]{acronym} 
\usepackage{color,soul}
\newcommand{\red}[1]{\textcolor{black}{#1}}

\newcommand{\droneposition}{{\it'UAV stop-point'} }
\newcommand{\dronepositions}{{\it'UAV stop-points'} }

\begin{document}

\title{Understanding UAV-Based  \red{{WPCN}}-Aided Capabilities for Offshore Monitoring Applications}

\author{
Dick Carrillo$^{1}$, Konstantin Mikhaylov$^{2}$, Pedro J. Nardelli$^{1,2}$, Sergey Andreev$^{3}$, and Daniel B. da Costa$^{4}$ \\
$^{1}$LUT School of Energy Systems, Lappeenranta-Lahti University of Technology, Finland\\
$^{2}$Centre for Wireless Communications, University of Oulu, Finland\\
$^{3}$Unit of Electrical Engineering, Tampere University, Finland \\
$^{4}$Department of Computer Engineering, Federal University of Cear\'{a}, Brazil\\
}

\maketitle
\vspace{-5mm}
\begin{abstract}

Despite the immense progress in the recent years, efficient solutions for monitoring remote areas are still missing today. This is especially notable in the context of versatile maritime and offshore use cases, owing to a broader span of operating regions and a lack of radio network infrastructures. In this article, we address the noted challenge by delivering a conceptual solution based on the convergence of \red{three emerging technologies -- unmanned aerial vehicles (UAVs), battery-less sensors, and wireless powered communication networks (WPCNs). Our contribution offers a systematic description of the ecosystem related to the proposed solution by identifying its key actors and design dimensions together with the relevant resources and performance metrics. A system-level modeling-based evaluation of an illustrative scenario delivers deeper insights into the considered operation and the associated trade-offs}. Further, unresolved challenges and perspective directions are underpinned for a subsequent study.
\end{abstract}

\input{acronyms.tex}
\section{Introduction}

\red{The technological developments around unmanned aerial vehicles (UAVs) are bringing notable advantages for the provisioning of important services.}
%
%
\red{Among them, one can cite remote monitoring and inspection, localization and tracking for search and rescue, and autonomous delivery of goods.}
%
\red{Even the most conservative industrial sectors} like  construction, production, and energy are already employing UAVs for remote structural monitoring and regular inspection operations \cite{uav_remotesensing_xiang}.

\begin{figure*}[t]
\centering
\includegraphics[scale=0.6]{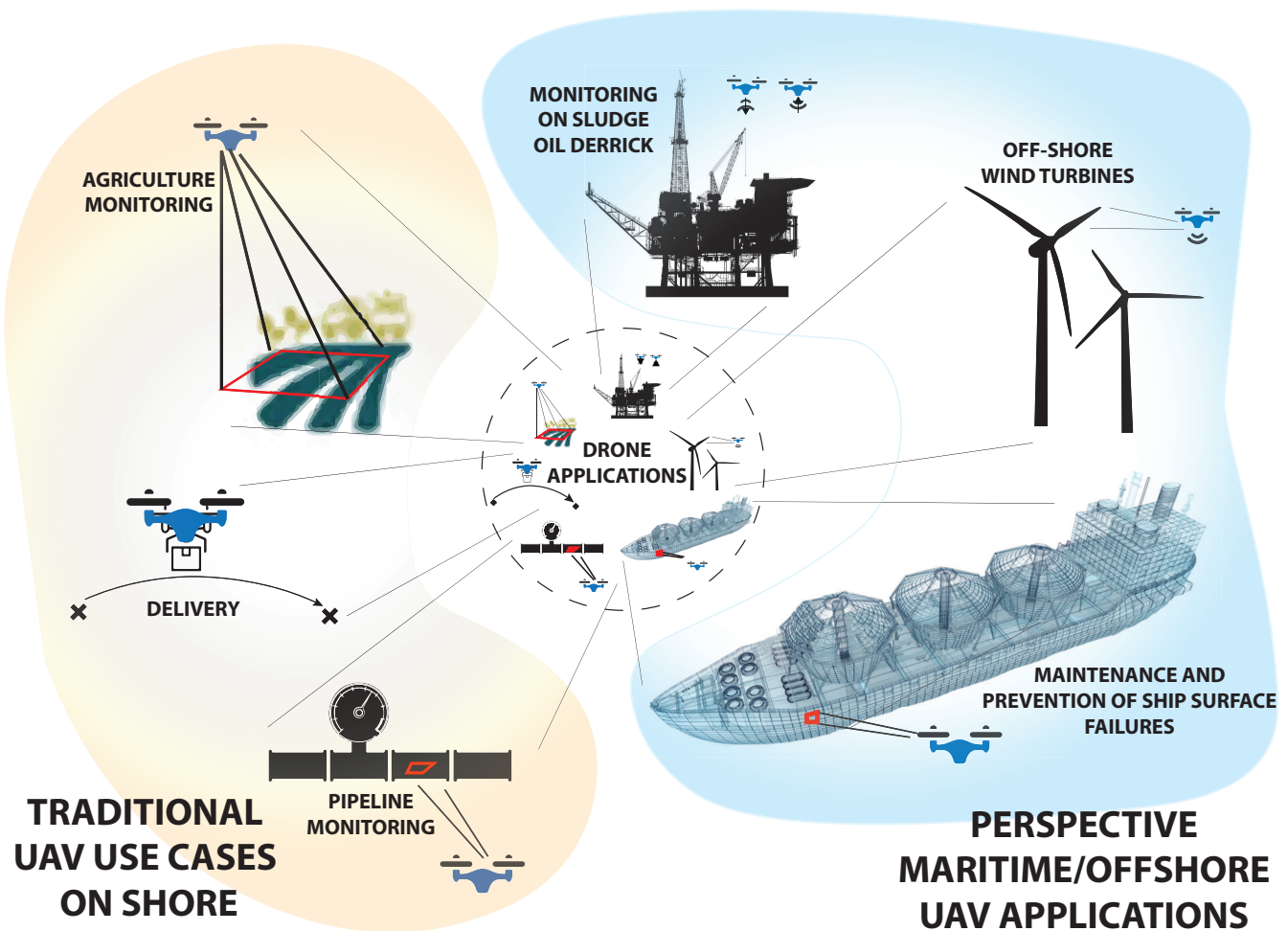}
\caption{Potential applications of UAV-based \ac{WPCN}-aided sensing.}
\label{fig:illustration_uavs}
\end{figure*}

%
\red{To offer a compelling example, UAVs carrying thermal cameras are employed to assess the structural integrity and condition of dams, agriculture monitoring, bridges, electricity grids, and pipelines; some of these use cases are described in Fig. \ref{fig:illustration_uavs}.} 
However, the data collected by drones are typically processed today by a human operator either online, where the operator may also act as a pilot, or in the offline regime. 
%
\red{To this end, the lack of autonomy and the limited support for unattended operation constitute fundamental challenges, thus obstructing further adoption of the UAV technology across sectors and verticals.} 
%
\red{One of the domains that embodies a plethora of use cases where the UAVs remain underutilized, and which eagerly anticipates the time when such a technology matures, is maritime and offshore operations~\cite{maritime_iot_xia}}

%
\red{In~\cite{uav_maritime_transvina}, the four key areas for drone applications in maritime contexts have been established, namely, data gathering, surveillance, rescue, and autonomous delivery.}
\red{The \ac{EMSA}~\cite{EMSA_usecases_ref}, in its turn, has identified the monitoring of pollution (e.g., in the air or water), surveillance (e.g., illegal fishing and drug traffic detection), patrol (e.g., tracking and identification of vessels), and rescue as the perspective scenarios to be served by the UAVs.}
\red{The main challenges obstructing efficient use of drones in the sea are: (i) limited operation time of today's UAVs, (ii) intermittent nature of the available maritime radio connectivity solutions, and (iii) insufficient sensory and computing capabilities of individual drones.}

%
\red{In this context, our contribution here is a comprehensive and systematic assessment of the key system design dimensions with their respective resources and actors that are employed to identify the trade-offs associated with the utilization of UAVs for maritime and offshore operations.}
%
\red{This discussion is accompanied by our analysis of an illustrative scenario, which we consider to be a perspective and, notably, a feasible candidate for leveraging the UAVs in the maritime context.}
\red{The targeted use case brings together several groundbreaking technologies, i.e., the UAVs, the \acp{WPCN}, and the airborne-deployed biodegradable sensors, to enable infrastructure-less, ecological, and robust collection of rich sensory data.}
\red{Its potential applications involve, but are not limited to, preemptive maintenance of vessels and offshore structures, pollution monitoring, safety, and rescue, among many others.}

The rest of this text is organized as follows.
Section II \red{outlines the setup and introduces the terminology employed in this work, as well as offers a description of the enabling technologies and their mechanisms}.
Section III reviews the landscape of \red{maritime UAV applications by discussing the relevant actors, their motivations, resources, and the involved trade-offs}.
Section IV elaborates on the essential system design dimensions and their interrelation.
Selected numerical results and further details on our illustrative scenario are presented in Section V. Finally, Section VI concludes the article by formulating the lessons learned and highlighting the perspective directions for further studies.

\section{\red{Convergence of UAV, Sensing, and WPCN to Enable Maritime Applications}}

\subsection{\red{Unmanned Aerial Vehicles}}
Throughout the last decade, UAVs or drones have been the focus of notable research and development activities,  \red{remarkably for monitoring, cargo delivery, and surveillance applications, as well as by operating as a communication platform}.
\red{To improve the drone autonomy, significant efforts were invested into increasing their awareness about the environment and enabling application-specific optimizations of the UAV trajectory~\cite{cognitivecomm_ullah}}.
%
\red{For instance, by aiming at a promising platform for delivering wireless connectivity, the optimization of spectral and energy efficiencies subject to data relaying, device-to-device communications, caching, mobile edge computing, and data collection has been introduced in \cite{UAV_wireless_optimization_yang2019}.}
%

\red{However, despite the immense technological progress, the active operation times for most of the contemporary commercial drones do not exceed several hours due to (i) drone's battery or fuel tank volumes, (ii) non-ubiquitous availability of radio connectivity, and (iii) control limits of an operator supervising the UAV.}
\red{For maritime and offshore missions, this significantly limits the potential area of drone's operation, especially considering the more limited possibilities for UAV re-charging at the sea.}
\red{This situation can be improved, e.g., by allowing a drone to harvest the energy for its operation during a mission (either in-flight or while floating in water), or by ensuring the presence of a dedicated zone (e.g., an offshore structure, a vessel, or a balloon/UAV) where a drone can replenish its energy.} 

\subsection{\red{Sensing Technologies}}
\red{Being already widespread on-shore, wireless sensors and IoT devices are not yet as ubiquitous in the sea.}
%
\red{Among the key reasons behind this are: (i) larger span and higher dynamics of the sea environment, (ii) electronics-unfriendly situation requiring special protection as well as hampering installation and service, and (iii) ecological and economic aspects related to in-sea deployment and operation of modern sensors.}
\red{To this day, the vast majority of sensors in maritime contexts have been deployed statically -- attached either to a vessel or to an element of the infrastructure (e.g., an oil platform), as well as implemented as an anchored buoy.} 

\red{However, the recent advancements in bio-degradable electronics enable the devices to decompose naturally without polluting the environment~\cite{biodegradable_Khalili}. They have the potential to change the above long-standing paradigm~\cite{iot_paper_sensors_fitzgerald, biodegradable_sensor_osaka,biodegradable_Khalili}. }
%
\red{Indeed, miniature and low-cost bio-degradable maritime sensors allow to support temporary mission-specific deployments, while at the same time enable these sensors to be carried by the sea waves.}
\red{The energy required by these sensors can be supplied either with a biodegradable battery, an energy harvesting element, or by channeling it remotely through wireless power transfer techniques.}

\subsection{\red{Wirelessly-Powered IoT}}
\red{Over-the-air wireless energy transfer is one of the few technologies that enables continuous operation of wireless sensors and IoT devices, by also stripping them of bulky, costly, and environmental unfriendly batteries.}
\red{Previous studies have indicated that \ac{WPCN} sensors powered via wireless power transfer by designated or opportunistic chargers can enable virtually unlimited lifetimes.}
%
\red{Deploying such a charger on a mobile platform as, for instance, a drone allows to increase the chances of having the line-of-sight operation between the charger and the sensor, thus maximizing the efficiency of energy transfer.}
\red{In \cite{swipt_baek2020}, a joint optimization of UAV hovering location and sensor energy consumption was pursued, where a geometry-based algorithm was employed to establish the best drone routes and hovering durations.}

\subsection{\red{Convergence and Application}}
\label{subsec:convergence}
\red{The above technologies may converge seamlessly in the maritime context by establishing a foundation for multiple novel applications, as sketched in Fig. \ref{fig:factors}. While being a single example, it serves as an illustrative case in terms of the existing perspectives and challenges.}
\red{Consider a vessel or an offshore installation (e.g., an oil platform) that hosts one or several drones.}
\red{To support the UAVs, a designated land-and-charge area might be instrumented on the vessel hull.}

\red{The drones can be equipped with infrared cameras and other relevant on-board sensors, as well as have a mechanism to deploy miniature biodegradable IoT sensors -- these measure, e.g., water pollution or tension of the structures to detect corrosion, cracks, and indents \cite{UAVs_usecases_ref} -- and a wireless power transmitter for powering these.}
\red{Whenever required, the drone(s) may take-off in the air, by operating either fully autonomously or under an operator's control, and robustly and efficiently deploy the sensors around the area of interest.}
\red{By revisiting the deployed sensors, the UAV(s) can collect their data and replenish the sensor energy through wireless power transfer mechanisms, thus allowing them to continue operation.}
Importantly, wireless power transfer between the UAV and the sensor allows to eliminate the battery of the latter, hence making its design simpler, cheaper, and more nature-friendly~\cite{EH_battery_Hassan}.


\label{sec:case_study}
\section{Actors, Resources, \red{and Trade-offs} in UAV-based  \red{\ac{WPCN}}-Aided Applications}

\red{Departing from the discussed illustrative scenario, we present here a more comprehensive and systematic assessment of the potential operation landscape.}
\red{To this aim, we identify the key actors and the resources as well as capture the related trade-offs and highlight the associated system design dimensions.}

\subsection{Main Actors}
In various maritime applications that involve UAVs, three ecosystem actors can be identified in addition to the drone itself, as described below.
\subsubsection{Sensing}
Sensor--actuator systems, with which the UAV communicates directly to obtain (or inject) the data of interest or the service information to support the completion of a mission (e.g., navigation and weather data, or sensors deployed by a harbor operator). 

\subsubsection{Connectivity}
Communication infrastructure that the UAVs and other discussed actors might employ to connect with external systems or between each other \red{(e.g., commercial on-shore infrastructure, satellite network, or other vessels)}.
Depending on the mission and the environment, the network infrastructure can be based on a suitable wireless technology --\red{from WiFi to satellite, and from proprietary to cellular}.
\subsubsection{Power supply}
Power infrastructure (e.g., a charging station for the drone(s), a wireless power transmitter, or the main electricity network of a vessel) is the third actor that pertains to the UAV and sensing operations. Its function is to replenish the energy \red{(or gas, in case of combustion engine driven UAVs)} by either charging or replacing the physical battery.

\red{Importantly, a certain physical element of the system may execute multiple different functions.}
\red{As an example, for the scenario sketched in subsection \ref{subsec:convergence}, the UAV may initially utilize its on-board cameras to locate the deployed sensors, then employ wireless power transfer to deliver them energy for making the measurements, and, finally, collect, process, and forward the data to the vessel.}
\red{In a similar fashion, the vessel or the offshore platform may supply energy to the drone (thus acting as an element of the power infrastructure), provide it with connectivity services, and issue the relevant data or commands.}
Note that depending on the scenario at hand, the power supply and the radio communication functions (or other services) can be made available either free of charge or at an additional cost.

%
\subsection{Essential Resources}
Following the above lines, data, energy, and monetary costs become the key resources involved into the considered scenario.

\subsubsection{Data} Collecting the relevant information is typically required to accomplish a certain UAV mission.
For certain use cases, telemetry data assembled by the UAV itself (e.g., recorded video or measured gas concentration) suffices, whereas for other scenarios, having additional input from the surface sensors may be desired.
\subsubsection{Energy} A drone consumes energy for whatever it does: hovering, sensing, data processing, and communication.
In the absence of a dedicated power infrastructure, sensors and actuators must be fueled by batteries or apply energy harvesting techniques, which may limit their energy budgets.
This situation becomes even more constrained if UAVs are used to charge sensors/actuators via the \red{\ac{WPCN}} technology.
Despite the immense progress in energy harvesting efficiency, achieving energy-neutral UAV operation remains close to impossible.

\subsubsection{Expenses} Monetary costs are always present as a crucial factor to be considered.
Together with UAV-specific and sensor--actuator network-centric capital expenditures and service costs, data communication and energy replenishment might incur further expenses, especially if a third party provides these services.
These expenditures may be particularly significant if data processing is handled by a remote cloud, which can be complemented by relying on the edge computing capabilities to improve the processing efficiency and reduce the costs~\cite{edge_commputing_park}.
\red{In the specific scenario of maritime applications, the costs may grow substantially. This is because the sensors can be located in places that are difficult to access or where the vessel's power supply is cumbersome due to physical installation constraints and/or architecture requirements.}

\subsection{\red{Trade-Offs}}
The above resources are deeply interdependent, and their specific management largely determines the system behavior and performance.
Regardless of the actual situations and parameters for sensing, acquiring extra data typically implies higher energy consumption.
Further, handling larger volumes of data also generates additional energy expenditures and may involve increased communication costs.
Ultimately, purchase of the required energy from the grid or deployment of drones equipped with \red{\ac{WPCN}} charging functions adds to the monthly bill of system operators.

\section{Key System Design Dimensions}

\begin{figure}[t]
\centering
\includegraphics[scale=0.5]{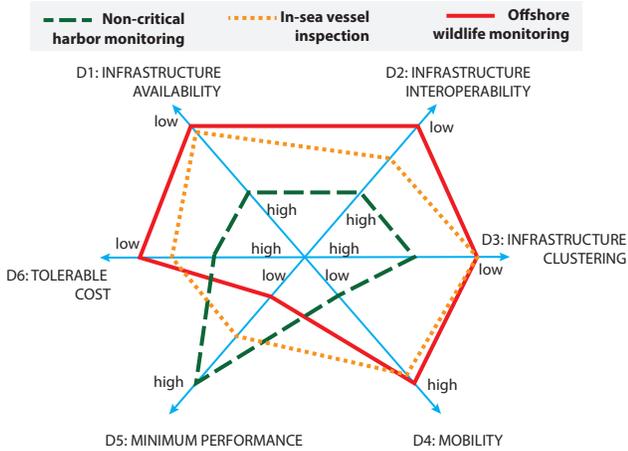}
\caption{Design dimensions and illustrative applications.}
\label{fig:factors}
\end{figure}

With UAVs becoming the central element of the scenario under study, six key design dimensions are identified in respect to the three ecosystem actors reviewed in the previous section.
In what follows, we outline these dimensions by formulating them as system design considerations. They are also grouped into three appropriate domains according to their relevance to the application development. To this end, Fig.~\ref{fig:factors} displays them across three illustrative maritime use cases.
%
%
\subsection{Feasibility}
The existence of the respective infrastructure (Dimension 1, D1) and the availability of the UAVs to utilize the offered services is the first and most crucial consideration. For example, if no infrastructure is available in a remote area, \red{the use of drones can become infeasible if they cannot operate autonomously. Another critical feasibility factor is the limited UAV operating times.}

Interoperability (D2) is an important issue that determines the feasibility of the discussed operation, even if the relevant infrastructure is deployed. The UAVs can locate and reach the said infrastructure, but it may be challenging to receive service when, for instance, the drone and the infrastructure are owned by different stakeholders or when their interfaces are not compatible.

\subsection{Dependability}
Assuming an existence and sufficient interoperability between the infrastructures, their topological characteristics (D3) can also affect the UAV operation.
For example, the optimal path of a drone collecting data and/or charging on-ground sensors differs depending on their spatial distribution (e.g., regular, clustered, or uniform).

The potential mobility (D4) of various infrastructure elements (e.g., sensors) and the respective temporal patterns (e.g., permanent vs. temporary deployments) is another crucial design factor.

\subsection{Efficiency}
The efficiency of service (D5) with respect to the key performance metrics (i.e., throughput, range, latency, and cost) is essential for the system design.
An inefficient \red{\ac{WPCN}} charging solution may serve as an example wherein the UAVs could spend most of their energy hovering to perform the \red{\ac{WPCN}} functions, while still being incapable of transferring sufficient amounts of energy to the sensors.

The tentative costs of service (D6) may become another limiting factor.
For instance, data transfer through a satellite link or a cellular connection while in roaming may incur substantial expenditures.

\red{It is important to note that the UAV's energy constraints impact the feasibility, dependability, and efficiency dimensions. As a consequence, the overall system performance is affected. To alleviate this impairment, various schemes arise as potential solutions to be considered, which includes trajectory design and renewable energy harvesting \cite{renewable_energy_uavs}.}

%
\section{Scenario and Modeling Results}
\red{To investigate the essential trade-offs associated with the considered system and our use case, we model the scenario sketched in subsection \ref{subsec:convergence}}.
\red{Specifically, we consider the situation of a vessel's integrity inspection as part of the preemptive maintenance procedure.}
\red{In what follows, we start by detailing our illustrative target scenario and modeling assumptions, and continue by discussing the representative results.}

\begin{figure*}[t]
\centering
\includegraphics[scale=0.6]{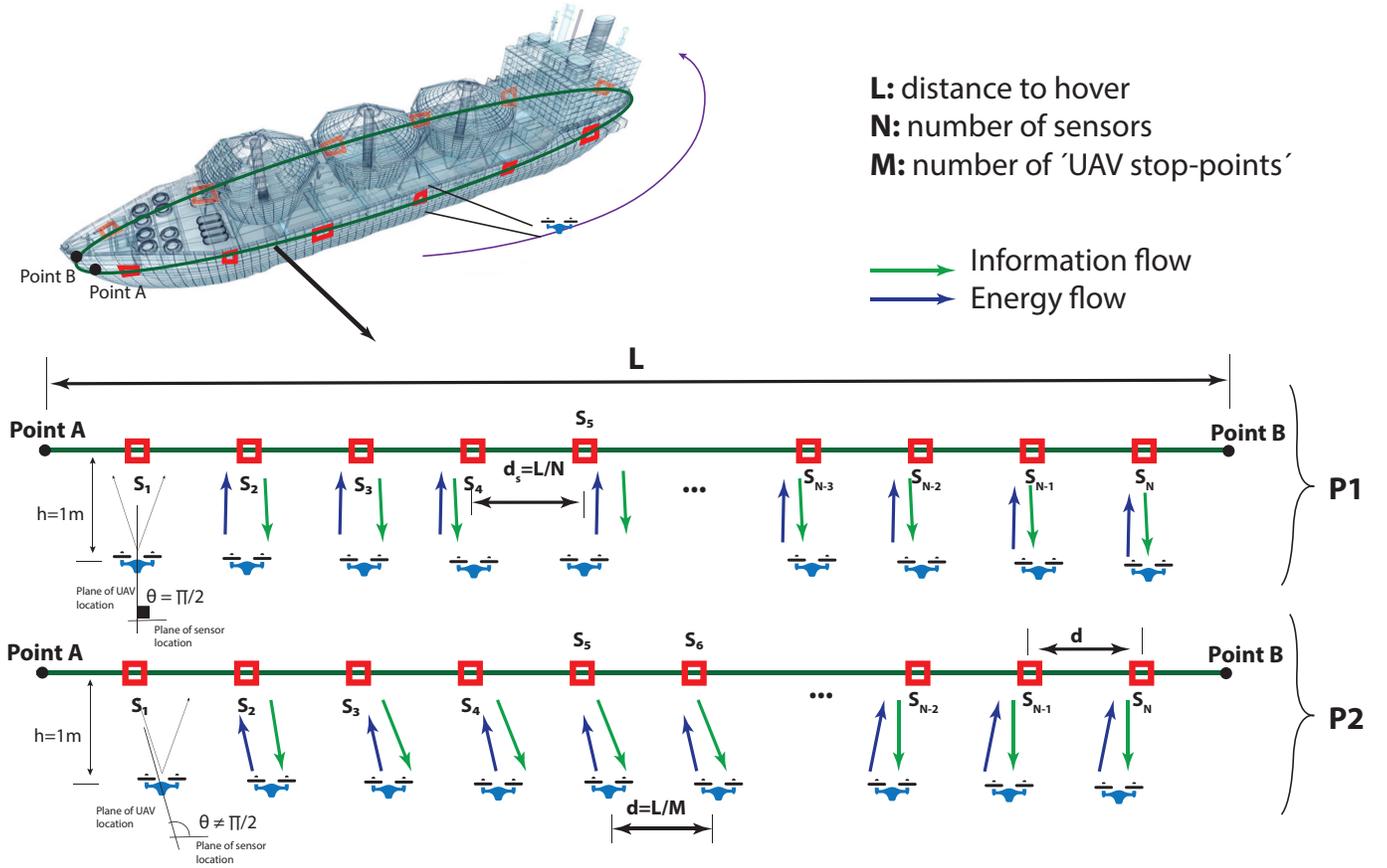}
\caption{Illustration of modeled scenarios.}
\label{fig:scenario}
\end{figure*}

\subsection{Scenario Description}
\red{Without loss of generality, we consider a \ac{WPCN} composed of (i) multiple biodegradable sensors deployed around a vessel at sea (modeled as an ellipse as depicted in Fig. \ref{fig:scenario}), which measure structural tension, and (ii) a single UAV that provides power for their operation.}
\red{The drone follows an ellipsoidal trajectory, while commuting between several static positions, which we refer to as \droneposition. It also hovers next to each of them for a certain period of time.}
\red{At each location, the UAV first employs wireless power transfer to charge the nearby sensors and then acquires the data generated by them}.

The efficiency of wireless power transfer is affected by the distance and the angle between the sensor and the charging drone.
Using the energy harvested from the \ac{RF} signal emitted by the UAV, the sensors make their measurements and send the data back to the charging drone (optionally, an additional \ac{WPCN} gateway may be deployed on the vessel to further increase the reliability of the system in question).
\red{The main parameters of our model, which have been aligned with the characteristics of the state-of-the-art commercial UAVs and sensors, are summarized in Table~\ref{table:sim_parameters}.}

\begin{table}[t!]
\centering
\caption{Model parameters (drone model is based on \cite{ref_real_development})}
\label{table:sim_parameters}
\begin{tabular}{|p{6cm}||p{1.4cm}| }
 \hline
 \hline

\textbf{Description} & \textbf{Value}\\
\hline
 Number of sensors  & 100 \\
\hline
 Number of \dronepositions  & 100/50/4 \\
\hline
\red{Flight path length}  & 500 m \\
\hline
\red{Distance from UAV flight trajectory to sensor} & 1 m \\
\hline
 Energy consumption of sensor for one measurement & 0.01 J\\
\hline
 Energy consumption of sensor for sending one packet & 0.01 J\\
\hline
\red{Energy consumption of UAV for receiving one packet} & 0.01 J\\
\hline
\red{Power consumption of drone in flight} & 170.3 W\\
\hline
\red{Drone/sensor power transfer antenna gain} & 9.3/8 dBi\\
\hline
\red{Distribution of time for energy delivery and data collection phases} & 50$\%$/50$\%$\\
\hline
\red{Efficiency of RF-DC conversion for sensor} & 72\% \\
\hline
 Minimum RF power enabling energy harvesting & 0 dBm \\
\hline
UAV battery capacity & 79,5 Wh \\
\hline
 Radio band & 2.3 GHz\\
\hline
\end{tabular}
\end{table}

\red{In our evaluation, we consider two different approaches to distribute the \dronepositions and sensors, as discussed in what follows.}

\subsubsection{\red{{Distribution of \dronepositions}}}
\begin{itemize}
\item Case P1: the UAV always stops in front of the sensor or a sensor cluster, which means that the angle between the plane where the sensors are placed and the direction toward the \droneposition is $\pi/2$ rad and the sum of the distances between the sensors and the drone is minimal. 
Clearly, this scenario implies the accurate knowledge of sensor positions by the UAV (or its ability to determine them).
\item Case P2: the UAV flight path is split by the stop points into a number of sectors having equal lengths. Note that in this case, the angle between the plane where the sensors are placed and the direction toward the UAV may differ from $\pi/2$.
This scenario does not require an exact knowledge of sensor locations.
\end{itemize}

\subsubsection{\red{Distribution of sensors}}
\begin{itemize}
    \item Case S1: the sensors are placed equidistantly across the entire measurement area.
    \item Case S2: the sensors are grouped, and their clusters are distributed equidistantly. For the sake of simplicity, a single cluster is composed of only two sensors.
\end{itemize}{}

\red{We note that our modeled scenario can be extended to many practical use cases in the context of maritime applications. For instance, if the UAV deploys the sensors, it can know the exact location of each battery-less sensor and use these data to plan its route most efficiently (case P1). The case P1 also characterizes the situation where a drone features a mechanism for active or passive sensor localization (e.g., by radio or machine vision). The alternative scenario (P2) reflects the case where the UAV cannot track the positions of sensors due to the absence of the respective technical means or when the sensor positions may change (e.g., if sensors are deployed in water).}

\red{The two sensor distribution scenarios reflect the extreme deployment strategies. Specifically, sensor clustering (S2) increases the reliability and efficiency of monitoring in dedicated areas (e.g., if the probability of a phenomenon occurrence or the value of data may vary across space). Further, a cluster of sensors may be deployed by the UAV to study a phenomenon more extensively. It can be detected by on-board sensing means (e.g., an infrared camera). Note that a cluster may comprise the sensors measuring different parameters. The more uniform distribution of sensors (S1) reduces the deployment costs whenever the values of data or the phenomenon occurrence probabilities have lower spatial dependency.}

\subsection{Essential Findings and Observations}

The described scenario was evaluated with our dedicated simulation environment built in Matlab, which is available open-source under \url{https://bitbucket.org/dcarrillom/uav_wpcn_maritime/}. 
The simulations are aimed to reflect the interrelations between the key system design dimensions and the trade-offs discussed in the previous section.

\begin{figure}[!t]
\centering
\includegraphics[scale=0.65]{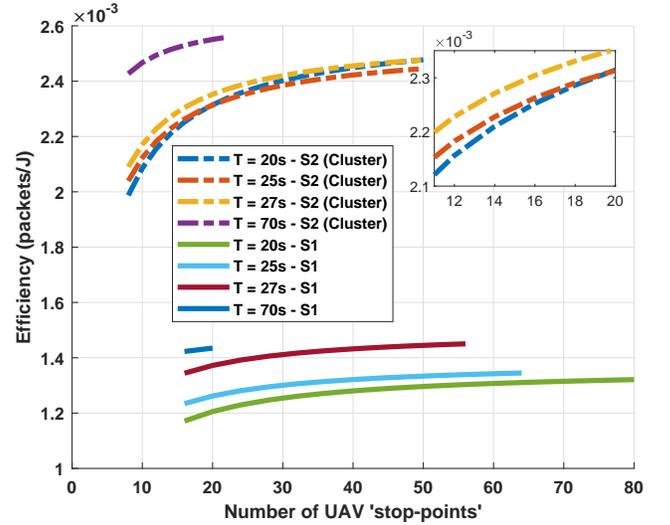}
\caption{Energy efficiency per packet vs. number of UAV \dronepositions (case P1, $T$ denotes the amount of time spent at each stop point).}
\label{fig:result1}
\end{figure}

\red{First, we focus on the \textbf{feasibility} of the target setup and the related technical limitations. Given the power/energy consumption and battery capacity figures in Table~\ref{table:sim_parameters}, and if residing in the air all the time, the maximum duration of a drone's mission approaches 28 minutes. Depending on the time spent at each \droneposition to charge the sensors and obtain data from them, the number of locations that the UAV can visit varies from a dozen to about eighty}. For instance, when spending at each \droneposition T=70 seconds, the UAV can visit 22 stops at most, as represented by the dashed purple line in Fig.~\ref{fig:result1}.

\red{Further, the \textbf{dependability} of our system is assessed by studying the possibility for the drone to visit all the sensors as well as the number of measurements that the UAV collects from each of them. Residing at each \droneposition for 20 seconds, the drone supplies the sensors with sufficient energy to carry and report five measurements, and can visit up to eighty various locations. An increase of the time spent at each \droneposition allows the sensors to obtain more energy and produce more packets, but reduces the number of different positions that can be served by the UAV.}

\begin{figure}[!t]
\centering
\includegraphics[scale=0.65]{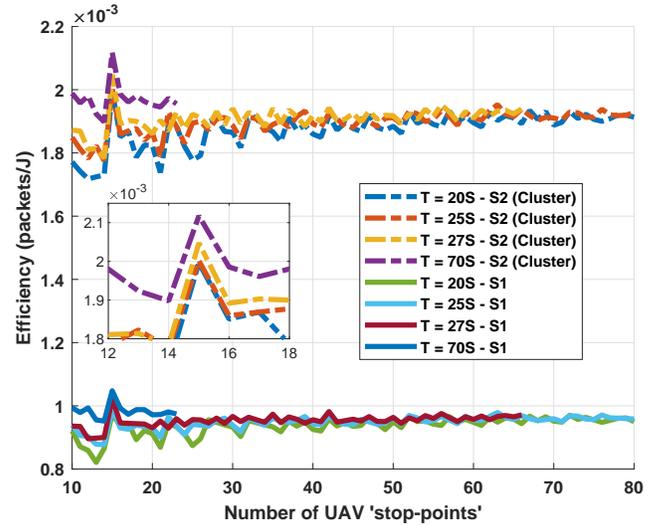}
\caption{Energy efficiency per packet vs. number of UAV \dronepositions (case P2, $T$ denotes the amount of time spent at each stop point).}
\label{fig:result2}
\end{figure}

\red{One of the crucial metrics related to the \textbf{efficiency} of the considered system is the number of packets transmitted by the sensors per a unit of energy consumed by the UAV. This parameter is reported in Figs.~\ref{fig:result1} and ~\ref{fig:result2} for the cases P1 and P2, respectively. The clustering of sensors (S2) substantially boosts these values by allowing multiple sensors to be charged simultaneously via wireless power transfer.}

\red{Comparing P1 and P2, one can observe that the knowledge of sensor locations enables 30\% improvement in the number of packets received per a unit of energy. An increase in the time spent by the UAV at each \droneposition further improves the energy efficiency. It is also worth noting that in Fig.~\ref{fig:result2} the efficiency reaches its peak for clustered and non-clustered deployments under different settings. This opens an opportunity to apply appropriate optimization techniques for the performance maximization in this or similar regimes.} 

%

\section{Conclusions and Research Challenges}

\red{
The maritime environment offers many challenges that the recent technological developments can help resolve. Occasionally, the best option is to design a dedicated system from scratch; at other times, this is infeasible or undesirable, since a combination of several existing techniques may become an efficient solution. In this work, we provided the initial insights into how a union of UAV and \ac{WPCN} technologies can enable dynamically-deployed infrastructure-less monitoring applications. Specifically, we first outlined the key actors, resources, and system design dimensions. Then we employed simulations to study the trade-offs within the illustrative scenario related to preemptive maintenance.}

\red{
Our numerical results suggested that there are several key parameters and considerations associated with the considered use case. They include the layout patterns of the sensors and the availability of accurate information on their location to the drone. This is especially relevant for the case where the sensors are deployed by the drone itself. For instance, sensor clustering allows to substantially increase the amounts of collected data (e.g., in our simulated scenario, the observed average energy efficiency growth for the sensor clusters was 1.7, i.e., 2.4/1.4). In contrast, a more uniform distribution of sensors is likely to produce a lower correlation between the measurements.}

\red{
It is also worth noting that different sensors can capture various parameters. In this case, cluster-based deployments may be particularly beneficial. Another crucial design dimension is the number and the position of the UAV stop points, and the amount of time spent by the drone at each of them to charge the sensors and to receive the data from them. For our modeled scenario, an increase in the stop time from 20 to 70 seconds brought a nearly 20\% improvement to the total volume of collected packets. However, this also reduced the number of UAV-visited locations from 80 down to 20. Importantly, our findings not only reported on the expected system performance but also revealed the existence of optimal operating points.}

\red{
As discussed in the first half of this article, the landscape of maritime UAV applications is broad and incorporates multiple potential actors. In the illustrative use case that we investigated, only several of them were considered. However, scenarios assuming the presence of other actors and instances deserve further research. Accordingly, several drones (i.e., a drone swarm) can operate across the scene. Another important use case addresses the (partial) availability of communication (e.g., data sinks) or energy (e.g., UAV or sensor chargers) related infrastructures. Finally, optimization of system operation under specific requirements (e.g., latency or event miss probability) represents another notable challenge. 
}
\section*{Acknowledgments}
This work is supported by the Academy of Finland via: (a) ee-IoT project n.319009, (b) FIREMAN consortium CHIST-ERA/n.326270, (c) EnergyNet Research Fellowship n.321265/n.328869, (d) project RADIANT (n.326196), (e) 6G Flagship (n.318927), (f) RoboMesh (n.336060), and (g) Project STREAM funded by Jane ja Aatos Erkon säätiö.

\bibliographystyle{IEEEtran}
\bibliography{ref_final}
\vspace{-12mm}
\begin{IEEEbiographynophoto}{Dick Carrillo}[S'01, M'06]
(dick.carrillo.melgarejo@lut.fi)
received the B.Sc.(2004) degree (Hons.) in Electronics and Electrical Engineering from San Marcos National University, Lima, Perú, and the M.Sc.(2008) degree in Electrical Engineering from Pontifical Catholic University of Rio de Janeiro, Rio de Janeiro, Brazil. Between 2008 and 2018, he participated on many research and development projects based on cellular network technologies such as WiMAX, LTE, LTE-A to support IoT Industrial applications. In 2019 he was a Visiting Research Fellow with the CTTC, Spain. Since 2018 he is a researcher at LUT University where he is also pursuing the Ph.D degree in Electrical Engineering. His research interests are mobile technologies beyond 5G, energy harvesting, reconfigurable intelligent surfaces, cell free massive MIMO, and grant-free access.
\end{IEEEbiographynophoto}
\vspace{-13mm}
\begin{IEEEbiographynophoto}{Konstantin Mikhaylov}[S’10-M’18-SM’19] (konstantin.mikhaylov@oulu.fi) is currently the Assistant Professor of Convergent Wireless for IoT with the CWC, University of Oulu, Finland. In 2019 he was a Visiting Research Fellow with the BUT, Czech Republic. He received Dr. Tech. degree from the University of Oulu in 2018, and M.Sc.(2008) and B.Sc.(2006) from the Saint Petersburg State Polytechnical University, Russia. He has (co-)authored over 90 published peer-reviewed papers on MTC and wireless connectivity for IoT, IoT devices and systems design, and applications. His research interests span over mMTC and URLLC IoT-grade wireless technologies, UAVs and ultra-energy-efficient IoT devices and systems design.
\end{IEEEbiographynophoto}
\vspace{-13mm}
\begin{IEEEbiographynophoto}{Pedro H. J. Nardelli}[M’07, SM’19] (pedro.nardelli@lut.fi) received the B.S. and M.Sc. degrees in electrical engineering from the State University of Campinas, Brazil, in 2006 and 2008, respectively. In 2013, he received his doctoral degree from University of Oulu, Finland, and State University of Campinas following a dual degree agreement. He is currently Associate Professor (tenure track) in IoT in Energy Systems at LUT University, Finland, and holds a position of Academy of Finland Research Fellow with a project called Building the Energy Internet as a large-scale IoT-based cyber-physical system that manages the energy inventory of distribution grids as discretized packets via machine-type communications (EnergyNet). He leads the Cyber-Physical Systems Group at LUT, and is Project Coordinator of the CHIST-ERA European consortium Framework for the Identification of Rare Events via Machine Learning and IoT Networks (FIREMAN) and of the project Swarming Technology for Reliable and Energy-aware Aerial Missions (STREAM) supported by Jane and Aatos Erkko Foundation. He is also Docent at University of Oulu in the topic of “communications strategies and information processing in energy systems”. His research focuses on wireless communications, energy management and cyber-physical systems.
\end{IEEEbiographynophoto}
\vspace{-13mm}
\begin{IEEEbiographynophoto}
{Sergey Andreev} (sergey.andreev@tuni.fi) is an associate professor of communications engineering and Academy Research Fellow at Tampere University, Finland. He has been a Visiting Senior Research Fellow with King's College London, UK (2018-20) and a Visiting Postdoc with University of California, Los Angeles, US (2016-17). He received his Ph.D. (2012) from TUT as well as his Specialist (2006), Cand.Sc. (2009), and Dr.Habil. (2019) degrees from SUAI. He (co-)authored more than 200 published research works on intelligent IoT, mobile communications, and heterogeneous networking.
\end{IEEEbiographynophoto}
\vspace{-13mm}
\begin{IEEEbiographynophoto}{Daniel B. da Costa}[SM] (danielbcosta@ieee.org) was born in Fortaleza, Ceará, Brazil, in 1981. He received the B.Sc. degree in Telecommunications from the Military Institute of Engineering (IME), Rio de Janeiro, Brazil, in 2003, and the M.Sc. and Ph.D. degrees in Electrical Engineering, Area: Telecommunications, from the University of Campinas, SP, Brazil, in 2006 and 2008, respectively. His Ph.D thesis was awarded the Best Ph.D. Thesis in Electrical Engineering by the Brazilian Ministry of Education (CAPES) at the 2009 CAPES Thesis Contest. From 2008 to 2009, he was a Postdoctoral Research Fellow with INRS-EMT, University of Quebec, Montreal, QC, Canada. Since 2010, he has been with the Federal University of Ceará, where he is currently an Associate Professor. In 2019, Prof. da Costa was on sabbatical leave. From January 2019 to April 2019, he was Visiting Professor at Lappeenranta University of Technology (LUT), Finland, with financial support from Nokia Foundation. He was awarded with the prestigious grant for Nokia Visiting Professor. From May 2019 to August 2019, he was with King Abdullah University of Science and Technology (KAUST), Saudi Arabia, as a Visiting Faculty, and from September 2019 to November 2019, he was a Visiting Researcher at Istanbul Medipol University, Turkey. 
\end{IEEEbiographynophoto}
\end{document}

%% file: acronyms.tex
\begin{acronym}
\acro{UAV}{unmanned aerial vehicle}
\acro{EMSA}{European maritime safety agency}
\acro{WPT}{wireless power transfer}
\acro{3GPP}{3rd Generation Partnership Project}
\acro{5G}{fifth generation of cellular networks}
\acro{RF}{radio-frequency}
\acro{3D}{three-dimensional}
\acro{CPS}{cyber-physical systems}
\acro{IoT}{Internet-of-Things}
\acro{M2M}{machine-to-machine}
\acro{LOS}{line-of-sight}
\acro{SWIPT}{simultaneous wireless information and power transfer}
\acro{WPCN}{wireless powered communication network}
\end{acronym} 